\documentclass[preprint2]{aastex}

\newcommand{\radm}{rad~m$^{-2}$}

\shorttitle{Enhanced Faraday rotation in spiral arms}
\shortauthors{Haverkorn et al.}

\begin{document}

\title{Enhanced small-scale Faraday rotation in the
   Galactic spiral arms} 

\author{M. Haverkorn\footnote{Harvard-Smithsonian Center for
    Astrophysics, 60 Garden Street, Cambridge MA 01238; mhaverkorn,
    bgaensler, nbizunok@cfa.harvard.edu} \footnote{Current address:
    Astronomy Department UC-Berkeley, 601 Campbell Hall, Berkeley, CA
    94720}, 
    B. M. Gaensler\footnotemark[1] \footnote{Alfred P. Sloan Research
    Fellow},
    J. C. Brown\footnote{Department of Physics and Astronomy,
    University of Calgary, 2500 University Drive N.W., Calgary, AB,
    Canada; jocat@ras.ucalgary.ca}, 
    N. S. Bizunok\footnotemark[1],
    N. M. McClure-Griffiths\footnote{Australia Telescope National
    Facility, CSIRO, PO Box 76, Epping, NSW 1710, Australia;
    naomi.mcclure-griffiths@csiro.au}, 
    J. M. Dickey\footnote{Physics Department, University of Tasmania,
    Private Bag 21, Hobart TAS 7001, Australia;
    john.dickey@utas.edu.au}, 
    A. J. Green\footnote{School of Physics, University of Sydney, NSW
    2006, Australia; agreen@physics.usyd.edu.au}} 

\begin{abstract}
We present an analysis of the rotation measures (RMs) of polarized
extragalactic point sources in the Southern Galactic Plane Survey.
This work demonstrates that the statistics of fluctuations in RM
differ for the spiral arms and the interarm regions. Structure 
functions of RM are flat in the spiral arms, while they increase in
the interarms. This indicates that there are no correlated RM
fluctuations in the magneto-ionized interstellar medium in the
spiral arms on scales larger than $ \sim 0^{\circ}.5$, corresponding
to $\sim 17$~pc in the nearest spiral arm probed. The non-zero slopes
in interarm regions imply a much larger scale of RM fluctuations. We
conclude that fluctuations in the magneto-ionic medium in the Milky
Way spiral arms are not dominated by the mainly supernova-driven
turbulent cascade in the global ISM but are probably due to a
different source, most likely H~{\sc ii} regions. 
\end{abstract}


\section{Introduction}

Structure in the neutral and ionized interstellar gas of the Milky Way
is ubiquitous and present on many scales. There have been many
observational, theoretical, and 
computational papers concerning turbulence in the neutral gas and
molecular clouds, but relatively little is known about the structure
of the warm ionized gas component (see Elmegreen \& Scalo 2004, Scalo
\& Elmegreen 2004 for an overview). Turbulence in the ionized gas is
suggested by observations of non-thermal linewidths in H$\alpha$
\citep{r85, trh99}, and modeled in numerical simulations of the
multiphase ISM (see V\'azquez-Semadeni et al. (2003) and references
therein). Furthermore, observed power spectra or
structure functions of electron density show power law behavior
indicative of incompressible hydrodynamical turbulence (Cordes et al.\
1985, Armstrong et al.\ 1995). However, care should be
taken in interpretation of these density spectra, as the connection
between density and velocity structure is not unambiguous, and
fluctuations in electron density can be created by a number of
different processes such as small-amplitude plasma waves or
differences in ionization fraction.

The observed density fluctuations in the ionized ISM in the Galactic plane
differ from those in the halo. The plane shows enhanced scattering of
extragalactic radio sources (Spangler \& Reynolds 1990), 
higher rotation measures (RMs) from extragalactic sources (e.g.\ Clegg
et al.\ 1992), and increased scintillation of pulsars and angular
broadening of extragalactic sources (e.g.\ Cordes et al.\
1985). Although fluctuations on scales of hundreds of parsecs have
been observed out of the  Galactic plane (e.g.\ Armstrong et al.\
1995), the gas in the plane may be partly dominated by structures on
much smaller scales (Haverkorn et al.\ 2004). Internal structure in
individual H~{\sc ii} regions in the Galactic plane may be responsible
for these enhanced fluctuations in the ionized ISM (Spangler \&
Reynolds 1990, Haverkorn et al.\ 2004). 

These previous studies have considered how the properties of
fluctuations in the ISM vary with Galactic latitude. In this {\em
Letter}, we examine structure in the thermal electron density and
magnetic field in the warm ionized ISM as a function of longitude, to
detect any change in characteristics between spiral arms and interarm
regions.

\section{The Southern Galactic Plane Survey}

The SGPS is a radio survey in the H~{\sc i} line and in 1.4~GHz
polarized continuum covering the ranges $253^{\circ}<l<358^{\circ}$ and
$5^{\circ}<l<20^{\circ}$ and $|b| < 1.5^{\circ}$ at $\sim1^{\prime}$
resolution, observed with the Australia Telescope Compact Array (ATCA)
and the Parkes 64m radio telescope (McClure-Griffiths et al.\ 2005;
Haverkorn et al. 2006). The polarized continuum has been observed only
with the ATCA, in twelve 8~MHz-wide frequency bands from 1336~MHz to
1432~MHz (where the 1408~MHz band is omitted due to radio
interference). This enables determination of RMs both from diffuse
Galactic synchrotron emission (Gaensler et al.\ 2001), and from
unresolved polarized extragalactic point sources, discussed here. RMs
of polarized point sources in the range $253^{\circ} < l <
358^{\circ}$ have been determined using the
technique explained in Brown et al.\ (2003); a full analysis of these
sources will be given elsewhere. RMs were determined for 151 sources,
the distribution of which can be approximated by a Gaussian centered
on +60~\radm\ with a standard deviation of 233~\radm.

\section{Structure functions of rotation measure}

The second order structure function, SF$_f$, of a function $f$
as a function of scale $r$ is defined as SF$_f (r) \equiv \langle [f(x) -
  f(x+r)]^2 \rangle_x$, where $\langle \rangle_x$ denotes averaging
over all positions $x$. We determined SFs of RM for five regions
at different Galactic longitudes, selected so that the major part of 
the probed sight line traced either spiral arms or interarm
regions. SFs were averaged azimuthally to obtain a 1D SF
as a function of angular scale $\theta$. Because a large-scale
gradient in the field will show up as a positive slope in the SF,
care has to be taken to correct for any large-scale gradients. Visual
inspection of the data did not show any clear gradients, and a first
order correction was made by subtracting a best-fit gradient from the
RM data before evaluating the structure function. 
The displayed errors are propagated errors in the RM values. These
errors are derived from the linear fit of polarization angle with
wavelength squared, and based on signal-to-noise ratios in Stokes $Q$
and $U$. Assuming the distribution is homogeneous and isotropic, the
effect of noise in the structure function can be corrected for. If the
observed RM is $RM_{obs} = RM(x) + \delta(x)$, where $\delta(x)$ is
the noise contribution, the SF of RM corrected for noise is $SF_{RM} =
SF_{RM_{obs}} - SF_{\delta}$ (Haverkorn et al.\ 2004).

\begin{figure}[t]
\epsscale{0.98}
\plotone{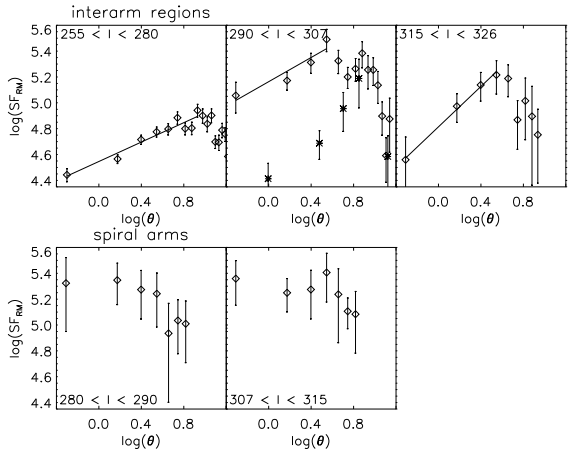}
\caption{Structure functions (SF) of RM as a function of angular
    scale $\theta$ in five regions of the inner Galactic plane
    (diamonds). For one region the SF of DM could be calculated
    (asterisks). The three upper plots show regions with sight lines
    predominantly through interarm regions, which show rising SFs. The
    bottom two plots sample mostly spiral arms, and show flat SFs,
    indicating that there is no correlated structure in the spiral
    arms on the scales sampled. The lines represent linear fits
    through the rising part of the SFs.} 
\label{f:sfrm}
\end{figure}

Fig.~\ref{f:sfrm} shows the SF of RM as a function of angular
scale $\theta$ for five regions in the SGPS. These were the regions
with the least confusion between spiral arms and interarm regions
along the line of sight. The top three plots, with
Galactic longitude range as denoted in the Figure, are at sight lines
predominantly through interarm regions, while the bottom two plots
show SFs of RMs mostly in spiral arms. The Galactic latitude range of
all regions is the latitude range of the SGPS, i.e.\ $|b| <
1.5^{\circ}$.  The behavior of the SF is different at the positions
of the spiral arms, indicating that the structure in RM changes in the
arms. In the interarm regions the SFs display a power law of RM
fluctuations, whereas the SF slopes in the spiral arms are consistent
with zero. The SFs in the interarm regions saturate at the maximum at
which flutuations exist, called the outer scale (or integral scale).

The SF slopes $\beta$ (SF$(r) \propto r^{\beta}$) for the three
interarm regions are $\beta=0.42\pm 0.04, 0.47\pm0.04$ and
$0.79\pm0.04$, with a weighted mean $\langle\beta\rangle =
0.55\pm0.05$. The probability that the spiral arm points are
consistent with $\langle\beta\rangle$ is $<10^{-10}$. The standard
deviation of RM seems to be higher in the spiral arms as well: the
standard deviation of RM in the interarm regions is $\sigma_{RM}=263\pm
32$~\radm, and the probability that the standard deviations of RM in
the selected spiral arm regions ($398\pm132$ and $415\pm125$~\radm\
respectively) are equal to $\sigma_{RM}$ is~0.07.

An independent determination of the fluctuations in thermal electron
density $n_e$ can be attempted using  H$\alpha$ observations to obtain
emission measures EM ($=\int n_e^2 ds$, where $ds$ is the path length
through the ionized gas). However, H$\alpha$ data suffer from high and
variable extinction in the Galactic plane, which introduces additional
power in the SF. Instead, we try an alternative method using dispersion
measures DM ($=\int n_e ds$) of  pulsars, which do not suffer from
extinction. Since pulsars are located inside the Galaxy, the path
length over which DM is calculated is smaller than the path length of
the RM measurements towards extragalactic sources. However,
since the approximate distances to the pulsars are known, we can
select a sample of pulsars sufficiently far away that their path
length is predominantly through spiral arms or through interarm
regions. Furthermore, we consider pulsars at approximately the same
distances, so that their spatial separation on the sky scales linearly
with their angular separation.

Pulsars were selected from the ATNF Pulsar
Catalogue\footnote{http://www.atnf.csiro.au/research/pulsar/psrcat}
(Manchester et al.\ 2005). We found 41 pulsars in the SGPS region with
distances $D$ in the range $8<D<10$~kpc using the DM values in the
catalog combined with the NE2001 model (Cordes \& Lazio 2003). Errors
in NE2001 are assumed to be mainly on small scales, so that the
uncertainty in the estimated pulsar distance is independent of spatial
scale. As a result, the error in the distance will increase the
amplitude of the SF, but not influence its scale.

Unfortunately the number of pulsars in each selected region was
so low that the SF of DM could be determined reliably only in the
region at $290^{\circ} < l < 307^{\circ}$, as shown in
Fig.~\ref{f:sfrm} as asterisks. Although inconclusive in itself, the
DM data confirm that SFs rise in an interarm region.

\section{Discussion and Conclusions}

We have shown that in the Galactic spiral arms no correlated
fluctuations exist in the magnetized interstellar plasma on scales
larger than $\sim0^{\circ}.5$, corresponding to $\sim 17$~pc in the
nearest spiral arm probed, i.e.\ the Carina arm which starts at
$\sim2$~kpc distance in this direction \citep{gg76, r03}. In the
interarm regions, however, correlated magnetoionic fluctuations {\it
  are} present on large scales. Since the RM is a line 
of sight average through the entire Galaxy, it is not possible to
associate a spatial scale to this angular scale. But assuming that the
largest angular scales represent nearby structure at a fairly
arbitrary distance of $\sim1$~kpc away, an outer scale of
$4-5^{\circ}$ in the interarm regions would correspond to a spatial
scale of about 100~pc.

The measured slopes of RM SFs in the interarm regions roughly agree
with slopes of {\it velocity} structure functions in simulations of
incompressible (magneto-) hydrodynamical turbulence \citep{k41, mg01,
clv02}. This suggests that the density and velocity spectra may be
coupled, which is only the case for subsonic or mildly supersonic
turbulence \citep{kr05, blc05}. Therefore, if the RM fluctuations are
connected to velocity structure, the turbulence in the interarm
ionized gas must be subsonic or only mildly supersonic, in agreement
with observational results from H$\alpha$.

On the other hand, the RM structure observed in spiral arms is
probably {\it not} a part of the turbulent cascade in the diffuse
ionized medium. If the density spectrum traces the velocity spectrum,
this would mean that the outer scale of turbulence would be $\la
17$~pc. However, the dominant source of turbulent driving is believed
to be supernova remnants and superbubbles, injecting energy on much
larger scales \citep{m03, db05}.

A shallow RM SF can also be caused by highly supersonic compressible
turbulence, which will flatten the density SF \citep{kr05,
blc05}. However, as mentioned before, it is doubtful whether the
ionized gas in the spiral arms is very supersonic.

Therefore, the fluctuations in RM in the spiral arms are probably not
connected to the turbulent cascade in the diffuse ionized ISM at
all. Instead, a probable source of these fluctuations is \ion{H}{2}
regions, which are of the correct size, sufficiently abundant
\citep{hgm04} and concentrated in spiral arms. Widespread interstellar
turbulence injected by \ion{H}{2} regions is not expected to be
significant \citep{m03}, but the structure could be caused by the
\ion{H}{2} regions themselves, or by turbulence inside them.

\acknowledgments
The authors would like to thank Mordecai Mac Low, Alex Lazarian and
Dick Crutcher for stimulating discussions, and the referee John Scalo
for valuable comments. The Australia Telescope is funded by the
Commonwealth of Australia for operation as a National Facility managed
by CSIRO. MH and BMG acknowledge the support of the National Science 
Foundation through grant AST-0307358.

\end{document}